\documentclass[a4paper,12pt]{article}
\usepackage{graphicx,epsfig}
\usepackage{epstopdf}
\usepackage{amssymb} 
\usepackage{amsmath}

\newcommand{\Eq}[1]{Eq.(\ref{#1})}
\newcommand{\rfn}[1]{(\ref{#1})}

\newcommand{\be}{\begin{equation}}
\newcommand{\ee}{\end{equation}}
\newcommand{\br}{\begin{eqnarray}}
\newcommand{\bea}{\begin{eqnarray}}
\newcommand{\eea}{\end{eqnarray}}
\newcommand{\er}{\end{eqnarray}}
\newcommand{\ba}{\begin{array}}
\newcommand{\ea}{\end{array}}
\newcommand{\bi}{\begin{itemize}}
\newcommand{\ei}{\end{itemize}}
\newcommand{\bn}{\begin{enumerate}}
\newcommand{\en}{\end{enumerate}}
\newcommand{\bc}{\begin{center}}
\newcommand{\ec}{\end{center}}

\def\gappeq{\mathrel{\rlap {\raise.5ex\hbox{$>$}}
{\lower.5ex\hbox{$\sim$}}}}

\def\lappeq{\mathrel{\rlap{\raise.5ex\hbox{$<$}}
{\lower.5ex\hbox{$\sim$}}}}

\begin{document}
\pagestyle{empty}
\begin{flushright}
\end{flushright}
\vspace*{15mm}
\begin{center}
{\large {\bf Testing neutrino masses in little Higgs models via discovery \\
of doubly charged Higgs at LHC}} \\
\vspace*{2cm}
{\bf A. Hektor}, {\bf M. Kadastik},  {\bf M. M\"untel},  {\bf M. Raidal}, and
 {\bf L. Rebane}
\vspace{0.3cm}

National Institute of Chemical Physics and Biophysics, Ravala 10,
Tallinn 10143, Estonia \\

\vspace*{3cm}
{\bf ABSTRACT} \\ 
\end{center}
\vspace*{5mm}
\noindent

We have investigated the possibility of direct tests of little Higgs models incorporating triplet 
Higgs neutrino mass mechanism at LHC experiments. We have performed Monte Carlo 
studies of Drell-Yan pair production of doubly charged Higgs 
boson $\Phi^{++}$ followed by its leptonic decays whose branching ratios are fixed from the
neutrino oscillation data.  We propose appropriate selection rules for the four-lepton 
signal, including reconstructed taus,  which are optimized for the discovery of $\Phi^{++}$ 
with the lowest LHC luminosity. As the Standard Model background can be 
effectively eliminated, an important aspect of our study is the correct statistical 
treatment of the LHC discovery potential. Adding detection efficiencies and measurement 
errors to the  Monte Carlo analyses, $\Phi^{++}$ can be discovered up to the mass 250 GeV 
in the first year of LHC, and 700 GeV mass is reachable for the integrated luminosity 
$L=30$ fb$^{-1}$.

\vspace*{2cm}
\noindent

\begin{flushleft} 
June 2007
\end{flushleft}
\vfill\eject

\setcounter{page}{1}
\pagestyle{plain}


\section{Introduction}

The main motivation of the Large Hadron Collider (LHC) experiment is to reveal the secrets of electroweak symmetry breaking. If the light standard model (SM) Higgs boson $H$ 
will be discovered, 
the question arises what stabilizes its mass against the Planck scale quadratically divergent 
radiative corrections. The canonical answer to this question is supersymmetry, predicting a very 
rich phenomenology of sparticles in the future collider experiments. 

Alternatively, the light SM Higgs boson may signal some strong dynamics at high
scale  $\Lambda\sim 4 \pi f,$ where $f$ is the decay constant of the new strongly 
interacting theory~\cite{Giudice:2007fh}. 
The most interesting class of models in such a scheme are the little Higgs models~\cite{Arkani-Hamed:2001ca,Cheng:2001vd,Arkani-Hamed:2001nc}. In those models the SM Higgs boson is a pseudo Goldstone mode of a broken global symmetry and remains much lighter than the other modes 
of the model, thus solving the little hierarchy problem and postponing the solution to the 
fundamental hierarchy problem to the scale $\Lambda.$ Those models are also very interesting
from collider physics point of view since they predict the existence of new particles, 
such as a new set of heavy  gauge bosons $W_H,\,Z_H$, a vectorlike  heavy quark pair $T, \, \bar T$ 
with charge 2/3, and triplet Higgs bosons $\Phi.$ 
If the new particle masses are ${\cal O}(1)$~TeV,
direct tests of the models are possible at 
LHC~\cite{Han:2003wu,Kilian:2003xt,Han:2005ru}.

An important open issue to address in the context of little Higgs models 
is the origin of non-zero neutrino 
masses~\cite{Lee:2005mb,Han:2005nk,delAguila:2005yi,Choudhury:2005jh,Abada:2005rt}.
The neutrino mass mechanism  which naturally occurs in those models is the 
triplet Higgs mechanism 
\cite{Schechter:1980gr,Ma:1998dx} 
(sometimes called type II seesaw) 
which employs a scalar with the $SU(2)_L\times U(1)_Y $ quantum 
numbers $\Phi\sim (3,2)$. 
The existence of such a multiplet in some versions of the little Higgs models 
is a direct consequence of global symmetry breaking which makes the SM Higgs 
light. For example, in  the minimal littlest Higgs model~\cite{Arkani-Hamed:2002qy}, 
the triplet Higgs with non-zero hypercharge arises from the breaking of global $SU(5)$ 
down to $SO(5)$ symmetry as one of the Goldstone bosons.
Its  mass $M_\Phi\sim g_s f,$ where $g_s<4 \pi$ is a model dependent coupling constant
in the weak coupling regime~\cite{Giudice:2007fh},
is therefore predicted to be below the cut-off scale $\Lambda$, and could be within the mass
reach of LHC.  Although the triplet mass scale is ${\cal O}(1)$~TeV, 
the observed neutrino masses can be obtained naturally. Firstly, non-observation of
rare decays $\mu\to eee,$ $\mu\to e\gamma,$ $\tau\to\ell\ell\ell,$ where $\ell=e,\mu,$ 
implies that the triplet Higgs boson Yukawa couplings $Y_{ij}$ must be small, 
thus suppressing also the neutrino masses. Secondly, the vaccuum expectation value
(vev) of the neutral component of triplet $v_\Phi$ contributes at tree level to 
the SM oblique corrections, and is therefore severely constrained by precision data.  
There exist additional mechanisms which can explain the smallness of $v_\Phi$ 
in little Higgs models.   Since the smallness of $v_\Phi$ is the most natural explanation
of the smallness of neutrino masses in the little Higgs models,  we assume 
this to be the case in this work.

The aim of this paper is to study the possibility of direct tests of little Higgs models
{\it and} neutrino mass mechanisms at LHC experiments via  pair productions and subsequent
decays of triplet Higgs boson. We study the Drell-Yan pair production of doubly charged component
of the triplet~\cite{Gunion:1996pq,Gunion:1989in,Huitu:1996su,Muhlleitner:2003me,
Akeroyd:2005gt,Rommerskirchen:2007jv}
\bea
pp\to\Phi^{++}\Phi^{--},
\label{production}
\eea 
followed by the leptonic decays.  Notice that 
$(i)$ the production cross section does not depend on
any unknown model parameter but the mass of $\Phi^{++}$;
$(ii)$ smallness of $v_\Phi$ in this scenario,  following from the smallness of neutrino masses,
 implies that the decays  $\Phi^{++}\to W^+W^+$ are negligible,
 and we neglect this channel in the following analyses;
$(iii)$ the $\Phi^{++}$ leptonic decay branching fractions do not depend on the size 
of the Yukawa couplings but  only on their ratios which are known from 
neutrino oscillation experiments. In the triplet model the normally hierarchical
light neutrino masses predict
$BR(\Phi^{++} \to \mu^+ \mu^+)\approx BR(\Phi^{++} \to \tau^+ \tau^+) 
\approx BR(\Phi^{++} \to \mu^+ \tau^+)\approx 1/3.$ 
Therefore this scenario is predictive and testable at LHC experiments.

The production process \rfn{production} has been studied 
before in various theory papers. In this work we first carry out a pure
 Monte Carlo study of the signal and background processes in the environment 
 of LHC detectors. After that we improve our analyses by adding particle
 reconstruction  efficiencies and  Gaussian distortion functions for particle 
 momentas and $E_T^{miss}$. Those mimic the detector inefficiency effects
 at the Monte Carlo level.
We believe that those effects  help us to estimate the realistic mass 
reach of the LHC detectors to the process under study.

In our study the new results are the following. For the signal reconstruction
we use new criteria, such as equality of invariant masses of positively and 
negatively charged  leptons together with total $\Sigma p_T$ cut for all leptons, which allows us 
to achieve better reconstructions efficiencies compared to the standard cuts.
We also reconstruct tau lepton final states with more than one $\tau$, which has 
not done before in this context.  As all the SM background can be eliminated 
in the case of this process, correct statistical analyses of the results in the
limit of no background is an important aspect of our study.
For the discovery criteria we have used the Log-Likelihood Ratio (LLR) statistical 
method to demand $5\sigma$ discovery potential to be bigger than 95\% ($1-\text{CL}_{s+b} > 0.95$). 
Our results are optimized for the discovery of process \rfn{production} with the lowest possible
LHC luminosity. 
The pure Monte Carlo study shows that $\Phi^{++}$ up to the mass 300 GeV is 
reachable in the first year of LHC ($L=1$ fb$^{-1}$) and  $\Phi^{++}$ up to 
the mass 800 GeV is reachable for the luminosity $L=30$ fb$^{-1}.$ 
Including the Gaussian measurement errors in the Monte Carlo the corresponding 
mass reaches become 250 GeV and 700 GeV, respectively. The errors of our estimates 
of the required luminosity for discovery depend strongly 
on the size of statistical Monte Carlo sample of the background processes.

The paper is organized as follows. In Section 2 we present the collider phenomenology 
of triplet Higgs boson and relate collider observables to neutrino mass measurements.
In Section 3  we discuss the Monte Carlo produced signal and background processes. 
In Section 4 we present the details of reconstruction and analysis procedure and results.
Detector effects are discussed  in Section 5. Finally we conclude in Section 6.

\section{Neutrino masses and collider phenomenology} \label{Sec2}

In this work we consider little Higgs scenarios in which, due to the breaking of global symmetry 
protecting the SM Higgs boson mass, the spectrum of the model contains also a 
pseudo Goldstone boson with the $SU(2)_L\times U(1)_Y $ quantum numbers 
$\Phi\sim (3,2)$~\cite{Arkani-Hamed:2002qy,Chang:2003zn}.
Although $\Phi$ is predicted to be heavier than the SM Higgs boson, 
the little Higgs philosophy implies that its mass could be
${\cal O}(1)$~TeV~\cite{Giudice:2007fh}. 
Due to the specific quantum numbers the triplet Higgs boson
couples only to the left-chiral lepton doublets $L_i\sim (2,-1)$, $i=e, \mu, \tau,$
via the Yukawa interactions given by 
\begin{equation}
L=i\bar L^c_{i} \tau_2  Y^{ij} (\tau\cdot \Phi) L_{j} 
+ h.c. ,
\label{L}
\end{equation}
where $Y_{ij}$ are the Majorana Yukawa couplings. 
The interactions \rfn{L}
induce lepton flavour violating decays of charged leptons which have not been
observed. The most stringent constraint on the Yukawa couplings comes from the 
upper limit on the tree-level decay $\mu\to eee$ and is\footnote{In little Higgs models
with $T$-parity there exist additional sources of flavour violation from the mirror
fermion sector~\cite{Choudhury:2006sq,Blanke:2007db}.}
$Y_{ee}Y_{e\mu}<3\cdot 10^{-5} \mathrm{(M/TeV)^2}$~\cite{Huitu:1996su,Yue:2007kv}.
Experimental bounds on the 
tau Yukawa couplings are much less stringent. 
In our collider studies we take $Y_{\tau\tau}=0.01$ and 
rescale other Yukawa couplings accordingly. In particular, hierarchical
light neutrino masses imply $Y_{ee},Y_{e\mu}\ll Y_{\tau\tau}$ consistently with the 
direct experimental bounds.

According to \Eq{L}, the neutral component of the 
triplet Higgs boson $\Phi^0$ couples  to the left-handed neutrinos with the 
same strength as $\Phi^{++}$ couples to the charged leptons. If $\Phi^0$
acquires a vev $v_\Phi$, non-zero Majorana masses are generated for the 
left-handed neutrinos~\cite{Schechter:1980gr,Ma:1998dx}. 
Non-zero neutrino masses and mixing is presently the only experimentally verified 
signal of new physics beyond the SM. In the triplet neutrino mass mechanism
the neutrino masses are given by
\begin{equation}
(m_\nu)_{ij} = Y_{ij} v_\Phi.
\label{mnu}
\end{equation}
We assume that the smallness of neutrino masses is explained by the smallness of 
$v_\Phi.$ In a realistic scenario massless Majoron, the Goldstone boson of 
broken lepton number, must be avoided. This is achieved by
an explicit coupling of $\Phi$ to the 
SM Higgs doublet $H$ via $\mu \Phi^0 H^0 H^0$~\cite{Ma:1998dx}, where
$\mu$ has a dimension  of mass. 
If $\mu\sim M_\Phi,$ in the concept of seesaw~\cite{seesaw}  the smallness 
of neutrino masses is  attributed  to the very high scale of triplet 
mass $M_\Phi$ because $v_\Phi=\mu v^2/M_\Phi^2,$ 
where $v=174$~GeV. 
However, in the little Higgs models the triplet mass scale 
${\cal O}(1)$~TeV  alone cannot suppress $v_\Phi.$
Therefore in this model $\mu\ll M_\Phi$, which can be achieved,
for example, via shining of explicit lepton number violation 
from extra dimensions as shown in ref.~\cite{Ma:2000wp,Ma:2000xh},
or if the triplet is related to the Dark Energy of the Universe~\cite{Ma:2006mr,Sahu:2007uh}.
Models with additional (approximate) $T$-parity~\cite{Chang:2003zn} make
the smallness of $v_\Phi$ technically natural. However, if the $T$-parity
is exact, $v_\Phi$ must vanish. In this work we do not consider the naturalness
criteria and assume that the above described neutrino mass scenario is realized in nature.
In that case $Y v_\Phi\sim {\cal O}(0.1)$~eV while the Yukawa couplings $Y$ can be
on the order of charged lepton Yukawa couplings of the SM.
As a result, the branching ratio of the decay $\Phi\to WW$ is negligible.
We also remind that $v_\Phi$ contributes to the SM oblique corrections, 
and the precision data fit $\hat T<2\cdot 10^{-4}$~\cite{Marandella:2005wd} 
sets an upper bound $v_\Phi \leq 1.2$~GeV on that parameter.

Notice the particularly simple connection between the flavour structure of light neutrinos
and the Yukawa couplings of the triplet via \Eq{mnu}. Therefore, independently of the overall size 
of the Yukawa couplings, one can predict the leptonic branching ratios of the
triplet  from neutrino oscillations. 
For the normally hierarchical light neutrino masses neutrino data implies negligible
$\Phi$ branching fractions to electrons and  
$BR(\Phi^{++} \to \mu^+ \mu^+)\approx BR(\Phi^{++} \to \tau^+ \tau^+)
\approx BR(\Phi^{++} \to \mu^+ \tau^+)\approx 1/3.$ 
Those are the final state signatures predicted by the triplet neutrino mass mechanism 
for collider experiments.

At LHC $\Phi^{++}$ can be produced singly and in pairs.
The cross section of the single $\Phi^{++}$ production via the $WW$
fusion process~\cite{Huitu:1996su} $qq\to q'q' \Phi^{++}$ scales as
$\sim v_\Phi^2.$ In the context of the littlest Higgs model
this process, followed by the decays $\Phi^{++}\to W^+W^+,$
was studied  in ref.~\cite{Han:2003wu,Han:2005ru,Azuelos:2004dm}.
The detailed ATLAS simulation of this channel shows~\cite{Azuelos:2004dm}
that in order to observe an $1$~TeV $\Phi^{++},$ 
one must have $v_\Phi>29$~GeV. This is in conflict with the 
precision physics  bound 
$v_\Phi \leq 1.2$~GeV as well as with the neutrino 
data. Therefore the $WW$ fusion channel is not experimentally promising for
the discovery of  doubly charged Higgs.

On the other hand,  the Drell-Yan pair production process
 $pp\to \Phi^{++}\Phi^{--}$ is not suppressed by any small coupling and
 its cross section is known up to next to leading order~\cite{Muhlleitner:2003me} (possible 
 additional contributions from new physics such as $Z_H$ are strongly suppressed
 and we neglect those effects here). 
Followed by the lepton number violating decays $\Phi^{\pm\pm} \to \ell^\pm\ell^\pm$, 
this process allows to reconstruct  $\Phi^{\pm\pm}$ invariant mass
from the same charged
leptons rendering the SM background to be very small in the signal region.
If one also assumes, as we do in this work,  that neutrino masses come from 
the triplet Higgs interactions, one fixes the $\Phi^{\pm\pm}$ leptonic
branching ratios. This allows to test the triplet neutrino mass model at LHC.

\section{Monte Carlo simulation of the signal and backgrounds}

The production of the doubly-charged Higgs is implemented in the PYTHIA Monte Carlo generator \cite{Sjostrand:2000wi}. The final and initial state interactions and hadronization have been taken into account. We have used the CTEQ5L parton distribution functions.

In the following analysis the normal hierarchy of neutrino masses and a very small value of the lowest neutrino mass is assumed. Such a model predicts that Higgs decay into electrons can be neglected and that there are three dominant decay channels for  $\Phi^{++}$ with approximately equal branching ratios:
\begin{itemize}
\item $\Phi^{\pm \pm} \to \mu^{\pm} \mu^{\pm}$,
\item $\Phi^{\pm \pm} \to \mu^{\pm} \tau^{\pm}$,
\item $\Phi^{\pm \pm} \to \tau^{\pm} \tau^{\pm}$.
\end{itemize}

We have studied only pair production of doubly charged Higgs due to the reasons pointed out above. $\Phi^{\pm\pm}$ pair decay products can combine to five different $\tau$ and $\mu$ combinations: $4\mu$, $3\mu1\tau$, $2\mu2\tau$, $1\mu3\tau$ and $4\tau$. Before reaching the detector, $\tau$ decays into an $e$, $\mu$ or a hadronic jet (marked as $j$ below) with branching ratios of 0.18, 0.17 and 0.65, respectively \cite{Yao:2006px}. $\tau$ hadronic jets and $\mu$-s are well visible and reconstructible in detector. The reconstruction of an energetic $\tau$ from electron decay is sensitive to detector effects, involving sophisticated background processes \cite{PDTRv1:2006a1}. In the current analyses we will neglect this channel, which will cause 31\%  loss of the total signal. Such loss is still sufficiently low and can be considered acceptable. Table \ref{t1} gives the cross sections and the Monte Carlo generated event numbers in our study.

\begin{table}
\begin{center}
\begin{tabular}{|l|l|l|l|}
\hline\hline
Process & Total $\sigma$ & N of events & Corresponding \\ 
        & ($fb$)         & generated    & luminosity ($fb^{-1}$) \\
\hline
Signal & & & \\ 
$M_\Phi$=200 GeV  & 7.78E+01 & 1.00E+05 & 1.28E+03 \\
$M_\Phi$=500 GeV  & 1.99E+00 & 1.00E+05 & 5.03E+04 \\
$M_\Phi$=1000 GeV & 5.58E-02 & 1.00E+05 & 1.79E+06 \\
\hline
Background & & & \\
$pp \to t \bar t \to 4\ell$ & 8.84E+04 & 2.55E+07 & 2.88E+02 \\
$pp \to t \bar t$ Z         & 6.50E+02 & 1.50E+05 & 2.3E+02 \\
$pp \to ZZ$                 & 2.12E+02 & 1.00E+05 & 4.72E+02 \\
\hline\hline
\end{tabular}
\end{center}
\caption{Cross-sections, numbers of Monte Carlo generated events and the corresponding integrated luminosities of the generated events. For the signal events we have taken the branching ratios BR($\Phi^{\pm\pm} \to \mu^\pm \mu^\pm$) = BR($\Phi^{\pm\pm} \to \mu^\pm \tau^\pm$) = BR($\Phi^{\pm\pm} \to \tau^\pm \tau^\pm$) = $1/3$.}
\label{t1}
\end{table}

The signatures of $\Phi$ decay are very clean due to 
$(i)$ high transfer momentum of the decay products, 
$(ii)$ lepton number violation and $(iii)$ pair production of $\Phi$. 
The Standard Model particles are lighter than $\Phi$, so the  background $\mu$-s and $\tau$-s must have smaller transverse energy and they do not produce an invariant mass peak
in $\mu^+\mu^+,$ $\mu^+\tau^+,$ $\tau^+\tau^+,$ final states. 
The present lower bound for the invariant mass of $\Phi$ is set by Tevatron to $M_\Phi\ge 136$~GeV \cite{Acosta:2004uj,Abazov:2004au}. In our study, four-lepton background processes with 
reasonable cross-sections and high $p_T$ leptons arise from three Standard Model
processes
\begin{itemize}
\item $pp \to t \bar t$,
\item $pp \to t \bar t Z$,
\item $pp \to ZZ$.
\end{itemize}

PYTHIA was used to generate $t \bar t$ and $ZZ$ background  ($t \bar t$ is forced to decay to $WWb\bar b$ and W leptonically). The CTEQ5L parton distribution functions were used. 
CompHEP was used to generate the $Z t \bar t$ background via its PYTHIA 
interface \cite{Boos:2004kh,Belyaev:2000wn}. All the datasets were generated 
 in Baltic Grid. In addition to background processes shown in Table \ref{t1}, some other four-lepton background processes exist 
involving $b$-quarks in the final state (for example, $pp \to b \bar b$). 
As such processes are very soft, it is possible to use the effective tagging methods  
\cite{PDTRv1:2006a2} and totally eliminate this soft background \cite{PDTRv2:2006a1}. 
Also, we do not 
consider possible background processes from the physics beyond the Standard Model.

\section{Reconstruction and analysis of the Monte Carlo data} \label{Sec3}

To study the feasibility of detecting the signal over background, 
we have to work with five possible reconstruction channels according to the 
following final states.
\begin{itemize}
\item $\Phi^{++}\Phi^{--}\to 4\mu$: The cleanest and most simple channel.
\item $\Phi^{++}\Phi^{--}\to 3\mu1\tau$: The channel is easily reconstructable using an assumption that the neutrino originating from the $\tau$ decay is collinear with $\tau$-jet and gives majority to the missing transverse energy ($E_T^{miss}$).
\item $\Phi^{++}\Phi^{--}\to 2\mu2\tau$: The signature can be reconstructed using the same assumptions for both $\tau$-neutrinos. The whole $E_T^{miss}$ vector has to be used here, while in the previous channel only one component was needed.
\item $\Phi^{++}\Phi^{--}\to 1\mu3\tau$: The channel can be reconstructed theoretically relying on an additional requirement that the two Higgs bosons have equal invariant masses. However, the reconstruction is very sensitive to the experimental accuracy of $E_T^{miss}$ determination.
\item $\Phi^{++}\Phi^{--}\to 4\tau $: The channel can not be reconstructed (and triggered by the single muon trigger).
\end{itemize}

First, we apply general detector related cut-offs for the Monte Carlo generated data. Generated particles were reconstructed within the pseudorapidity region $|\eta|<2.4$ and with transverse momentum higher than $5$ GeV. These are the natural restrictions of the CMS and ATLAS detectors at the LHC. Only the pseudorapidity region $|\eta|<2.4$ is reachable for the detector and only the events with $p_T>5$~GeV are typically triggered. These restrictions suppress mainly the soft Standard Model background. The efficiency of lepton reconstruction and charge identification rate are very high,
we use the values  0.9 and 0.95, respectively \cite{CMS:1997ki}.

The invariant mass of two like-sign $\mu$-s and/or $\tau$-s are calculated using equation:
\be
(m_I^{\pm\pm})^2=(p_1^\pm + p_2^\pm)^2,
\label{eq1}
\ee
where $p_{1,2}$ is the $\mu$ or $\tau$ 4-momentum. Since the like-sign signal of $\mu$-s or $\tau$-s originate from a doubly charged Higgs boson, the invariant mass peak measures the mass of doubly charged Higgs, $m_I=M_\Phi$.
4-$\mu$ final state allows to obtain invariant masses directly from \Eq{eq1}. In channels involving one or several $\tau$-s, which are registered as $\tau$-jets or secondary $\mu$-s (marked as $\mu'$ below), the momenta of jets has to be corrected according to the equation system:

\begin{eqnarray}
\mathbf{p}_\tau^i &=& k^i \mathbf{p}_{jet}^i \label{eq2} ,\\
\mathbf{p}_{Tmiss} &=& \sum_i \mathbf{p}_{T\nu}^i \label{eq3} ,\\
M_{\Phi^{++}} &=& M_{\Phi^{--}} \label{eq4} ,
\end{eqnarray}

where $i$ counts $\tau$-s, $\mathbf{p}$ marks 3-momentum, $\mathbf{p}_{T\nu}$ is the vector of transverse momentum of the produced neutrinos, $\mathbf{p}_{Tmiss}$ is the vector of missing 
transverse momentum (measured by the detector) and $k_i > 1$ are positive constants. \Eq{eq2} describes the standard approximation that the the decay products of a heavily boosted $\tau$  are collinear \cite{CMS:1997ki}. \Eq{eq3} assumes missing transverse energy only to be comprised of neutrinos from $\tau$ decays. In general, it is not a high-handed simplification, because the other neutrinos in the event are much less energetic and the detector error of $E_T^{miss}$ is order of magnitude smaller \cite{PDTRv1:2006a3}. Using the first two formulas, it is possible to reconstruct up to two $\tau$-s per event. Additional requirement of \Eq{eq4} allows to reconstruct the third $\tau$ per pair event, although very low measurement errors are needed.

A significant fraction of $\tau$-s (0.18) decay into $\mu'$-s that cannot be distinguished from primary $\mu$-s in the detector. Still, if reconstructed invariant masses of $\Phi^{++}$ and $\Phi^{--}$ are considerably different,
we can suspect that one or several $\mu$-s originate form $\tau$ decays. In such case we can again use \Eq{eq2}-\eqref{eq4} to correct the 4-momenta of decay products. When only one secondary muon is present, $E_T^{miss}$ points into the same direction as its $p_T$. Otherwise $E_T^{miss}$ is a superposition of neutrino transverse momenta. Such correction tightens the invariant mass peak of the signal and does not produce any artificial background.

The occurrence probability of different reconstruction channels are presented in Table~\ref{t2}. The second column shows probabilities of Higgs decay to $N$ $\mu$-s  and $M$ $\tau$-s. Next columns describe the final state after $\tau$ decay to $\mu'$-s and/or jets. Different columns mark the number of secondary $\mu$-s and the rows designate $\tau$-jets in the detector recordings. The events having at least one $\tau \to e$ in a final state are omitted in  our analysis, as well as events with $M > 3$ or $N(\mu')>2$. The proportions of reconstructible signatures are marked in a bold-face. The table shows that 0-3 jet channels together with $\mu$ correction are almost equally important and overall reconstructible channels comprise $64\%$ of total events.

\begin{table}[h]
\begin{center}
\begin{tabular}{|l|l|l|l|l|l|l|}
\hline\hline
Decay channel 	&After Higgs	 & \multicolumn{5}{c|}{After $\tau$ decay ( $ x = jet_{\tau}$)} \\ \cline{3-7}
 		       		&decay ($x=\tau$)		 & 0 ($\tau\to\mu'$) & 1 ($\tau\to\mu'$) & 2 ($\tau\to\mu'$) & 3 ($\tau\to\mu'$) & 4 ($\tau\to\mu'$) \\
\hline
$2\Phi \to 4\mu $    	& 0.1111 		& \bf{0.1111}	& \bf{0.0377} 	& \bf{0.0107} 	& 0.0012 	& 0.0001 \\
$2\Phi \to 3\mu 1x $ 	& 0.2222	 	& \bf{0.1443}	& \bf{0.0736}	& \bf{0.0125}  	&  0.0014	&	\\
$2\Phi \to 2\mu 2x $ 	& 0.3333		& \bf{0.1407} 	& \bf{0.0478}	& {0.0054} 	& 	 	& \\
$2\Phi \to 1\mu 3x $	&0.2222		& \bf{0.0610}	& {0.0207}	&			&		&\\
$2\Phi \to 4x $		&0.1111		& 0.0198		&			&			&		&\\
\hline
Sum				&1.0			& \multicolumn{5}{c|}{$\mathbf{0.64}+0.05$} 				\\ \cline{3-7}
\hline\hline
\end{tabular}
\end{center}
\caption{Probabilities of all possible decay chains for $\Phi$ pairs in our scenario.
 $"x"$ in the table marks $\tau$ or, after $\tau$ decay, $\tau$-jet. The reconstructed signatures are marked in bold, the remaining signatures were not reconstructed. After omitting the channels that include $\tau\to e$ decay, $69\%$  of the total signal is left. In total  $64\%$ of the signal has been reconstructed.}
\label{t2}
\end{table}

A clear signal extraction from the Standard Model background can be achieved using a set of selection rules imposed on a reconstructed event in the following order.
\begin{itemize}
\item \emph{S1}: events with at least 2 positive and 2 negative muons or jets which have $|\eta| < 2.4$ and $p_T > 5$ GeV are selected.
\item \emph{S2}: $\sum p_T$ (scalar) sum of 2 most energetic positive and negative $\mu$-s or $\tau$-jets has to be bigger than a certain value (depending on Higgs mass).
\item \emph{S3}: Z-tagging -- if invariant mass of the pair of opposite charged $\mu$-s or $\tau$-jets is nearly equal to $Z$ mass (85-95 GeV), then the particles are eliminated from the analysis.
\item \emph{S4}: as $\Phi$-s are produced in pairs, the reconstructed invariant masses (in one event) have to be equal. We have used the condition
\be
0.8 < m_{I}^{++} / m_{I}^{--} < 1.2.
\label{eq5}
\ee
If the invariant masses satisfy the condition then we include them to the histogram, otherwise we 
suspect that some $\mu$-s may originate from $\tau$ decay, and make an attempt to 
find corrections to their momenta according to the method  described above.
\end{itemize}

\begin{figure}[h]
\begin{center}
\includegraphics[width=0.49\textwidth]{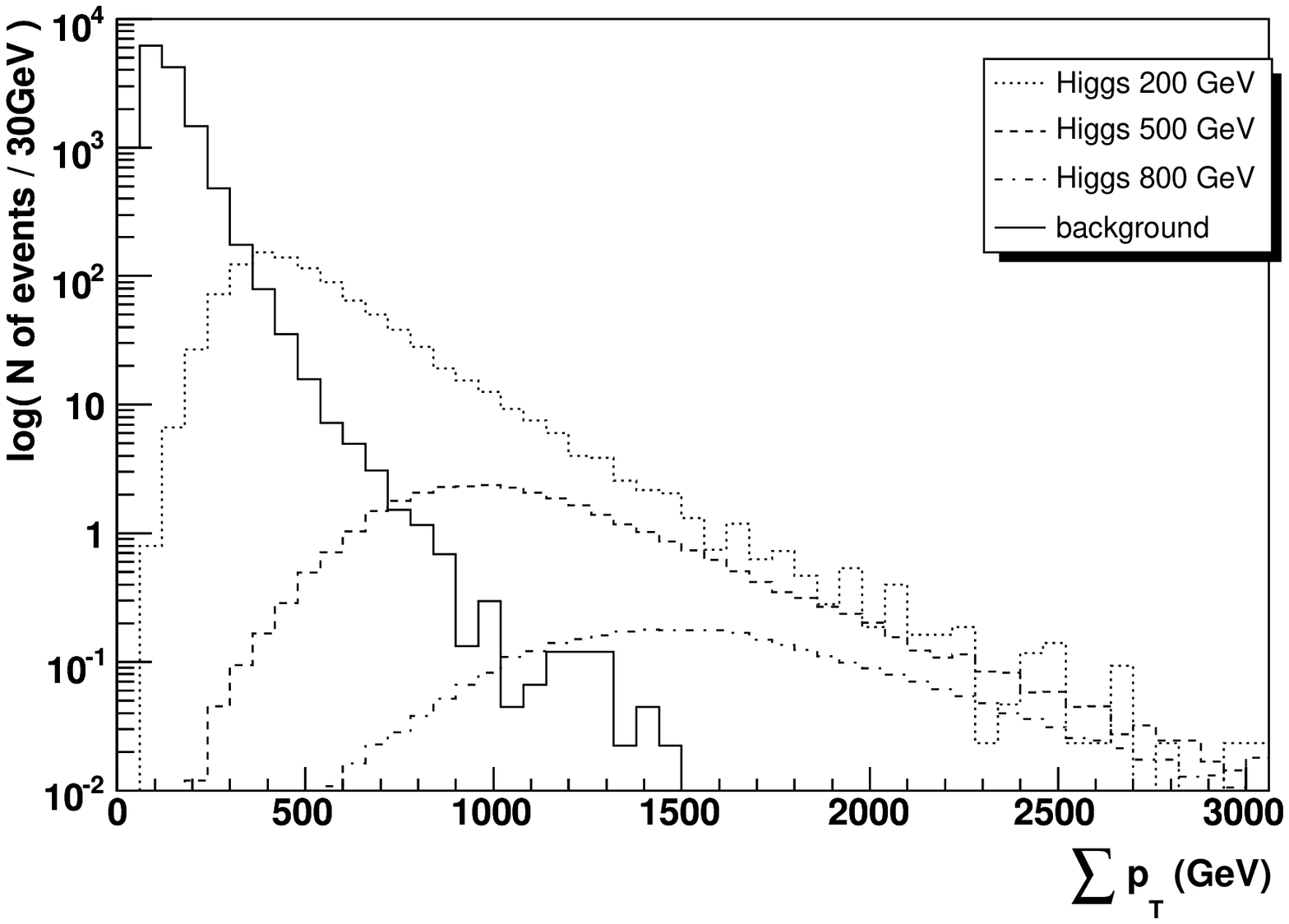}
\hfill
\includegraphics[width=0.49\textwidth]{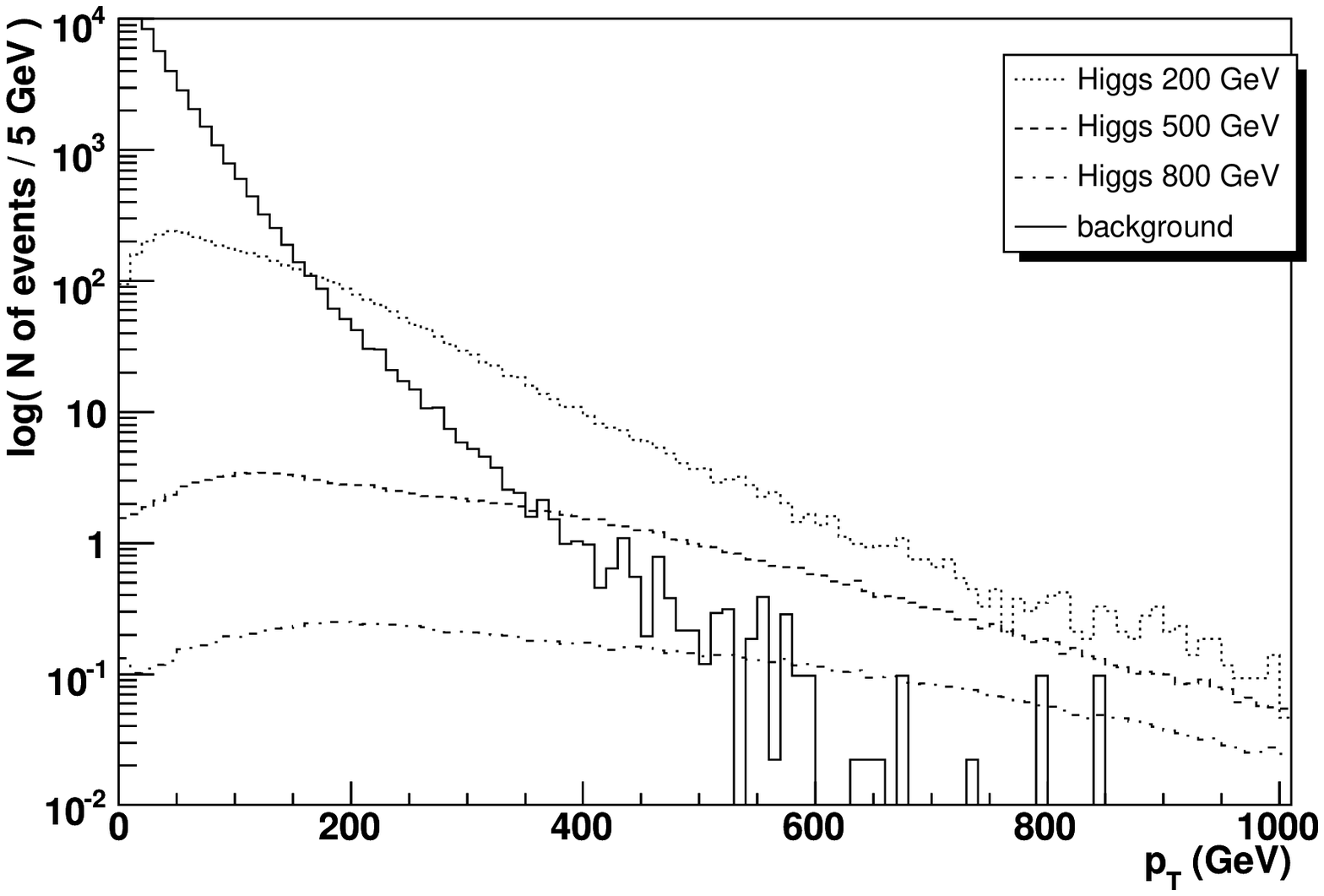}
\caption{The left panel shows the distribution of events according to scalar sum of 2 most energetic (highest $p_T$) positively and 2 most energetic negatively charged muons or jets ($\sum{p_T}$). The right panel shows the distribution of events considering traditional $p_T$ cut for single particles. Both figures correspond to luminosity $L = 30 fb^{-1}$.}
\label{f1}
\end{center}
\end{figure}

The rule S1 is an elementary detector trigger. S2, performing scalar sum of $p_T$, is an untraditional cut. The advantage compared to the widely used $p_T$ cut for a single particle is clearly visible from Figure~\ref{f1}. The left panel shows that the maximum of Higgs line reaches clearly out of the background while on the right panel the maximum is deeply inside the background.

\begin{figure}[h]
\begin{center}
\includegraphics[width=0.49\textwidth]{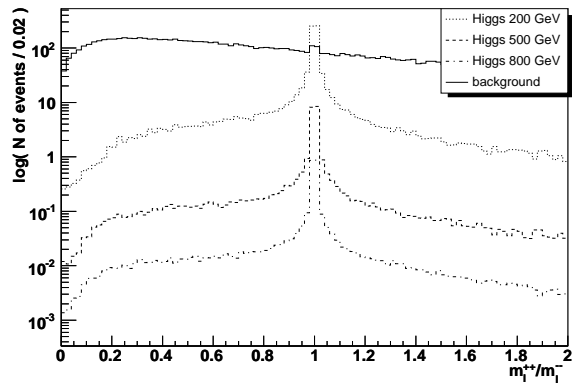}
\caption{Distribution of events according to the ratio of reconstructed invariant masses ($m_{\Phi^{++}} / m_{\Phi^{--}}$) (no other cuts are applied). The figure corresponds to luminosity $L = 30 fb^{-1}$.}
\label{f2}
\end{center}
\end{figure}

Z-tagging in S3 suppresses $pp \to ZZ$ and $pp \to t \bar t Z$ background. S4 is based on the equality of the invariant masses of like-signed $\mu$-s or $\tau$-s. Figure~\ref{f2} gives a clear picture of the behavior of signal and background for the S4 selection rule. Naturally, some freedom is needed due to the $\Phi$ decay width and experimental errors of the detector. We require that the ratio $m_{I}^{++} / m_{I}^{--}$ has to be in the region from 0.8 to 1.2. 

While the selection rules S1, S3 and S4 are independent of the Higgs mass,
the selection rule  S2 ($\sum{p_T}$ cut) has to be optimized for a certain Higgs mass value. 
The cut may be set to a very high value which eliminates all background events, but inevitable loss in signal may postpone the discovery of new physics at LHC. Thus it is natural to take the minimal discovery luminosity ($L_{min}$) as the optimization criteria. Looking for a cut value that enables to make a discovery with the lowest luminosity, we are dealing with small signal and background expectations by definition. Simple significance estimators cannot be exploited here. We have used the log-likelihood ratio (LLR) statistical 
method ~\cite{Read:2000, Barate:2003} to demand $5\sigma$ discovery potential to be bigger than $95\%$ ($1-\text{CL}_{s+b} > 0.95$) as for a discovery criteria. This is a rather strong requirement, because it allows to make a discovery (meaning the fluctuation of background may mimic the outcome of an experiment with probability less than $2.9\cdot 10^{-7}$ ($5\sigma$) ) during the specified luminosity with a probability of $95\%$ (if s+b hypothesis is correct). The widely used convention, that significance should exceed five, gives only $50\%$ discovery potential in Gaussian limit and diminishes to very small values when background approaches zero. 

The best value for S2 cut does depend on $M_\Phi$ but is not too sensitive to it.
Typically the $\sum{p_T}$ can be assigned a value with a precision of 100 GeV while affecting the minimum luminosity by only a couple of percent. In the Table~\ref{t4} the approximated middle point of this value is given.
As the best S2 cut is very strong, it suppresses almost entirely the generated background (being combined with the other selection rules). For Higgs masses above 500 GeV the background is totally suppressed and the discovery potential criteria meets the requirement for 3 signal events (6 invariant masses). Nevertheless we cannot infer that the background is really zero in nature. To estimate the statistical error due to final number of generated background events we have found 95$\%$ upper limit of background according to Poisson statistics (Table~\ref{t5}, in brackets). Using this limit in LLR analysis we get much higher luminosities for discovery. Even a very small background expectancy ($b = 0.01$) gives some possibility to have one ($9.9\cdot 10^{-3}$) or two ($4.9\cdot 10^{-5}$) background events in the experiment and these outcomes cannot be interpreted as discovery anymore. This phenomenon shifts the minimal required luminocity to much higher values denoted as $L_{max}$ in Table~\ref{t4}.

\begin{table}[h]
\begin{center}
\begin{tabular}{|l|l|l|l|l|l|l|l|l|l|}
\hline\hline
Mass of $\Phi$ (GeV) & 200 & 300 & 400 & 500 & 600 & 700 & 800 & 900 & 1000 \\
\hline
Optimal $\sum p_T$ for S2 (GeV)      & 300 & 400 & 600 & 700 & 860 & 860 & 860 & 860 & 860 \\
MC $L_{min}$ (fb$^{-1}$)  & 0.25 & 0.93 & 2.0 & 3.6 & 8 & 17 & 34 & 62 & 120 \\
MC $L_{max}$ (fb$^{-1}$)  &0.26 & 1.03 & 3.1 & 7.0 & 17 & 38 & 77 & 160 & 320 \\
\hline\hline
\end{tabular}
\end{center}
\caption{Optimal $\sum p_T$ cut for different Higgs masses and the corresponding 
minimal discovery luminosities: the lower ($L_{min}$) corresponds to the generated 
background in our analyses and the higher ($L_{max}$) corresponds to 95$\%$ upper limit of the background error.}
\label{t4}
\end{table}

\begin{figure}[h]
\begin{center}
\includegraphics[width=0.47\textwidth]{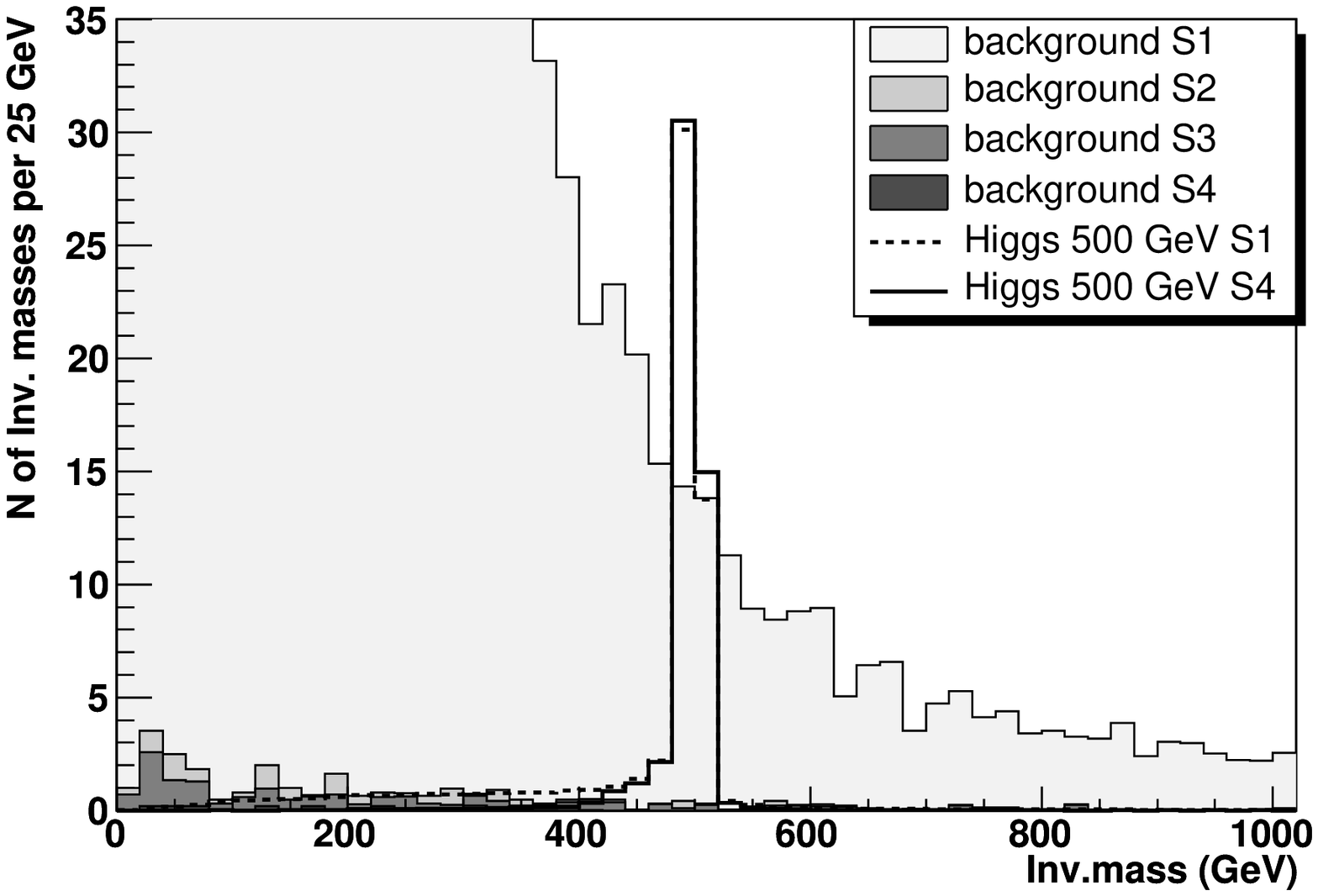}
\hfill
\includegraphics[width=0.47\textwidth]{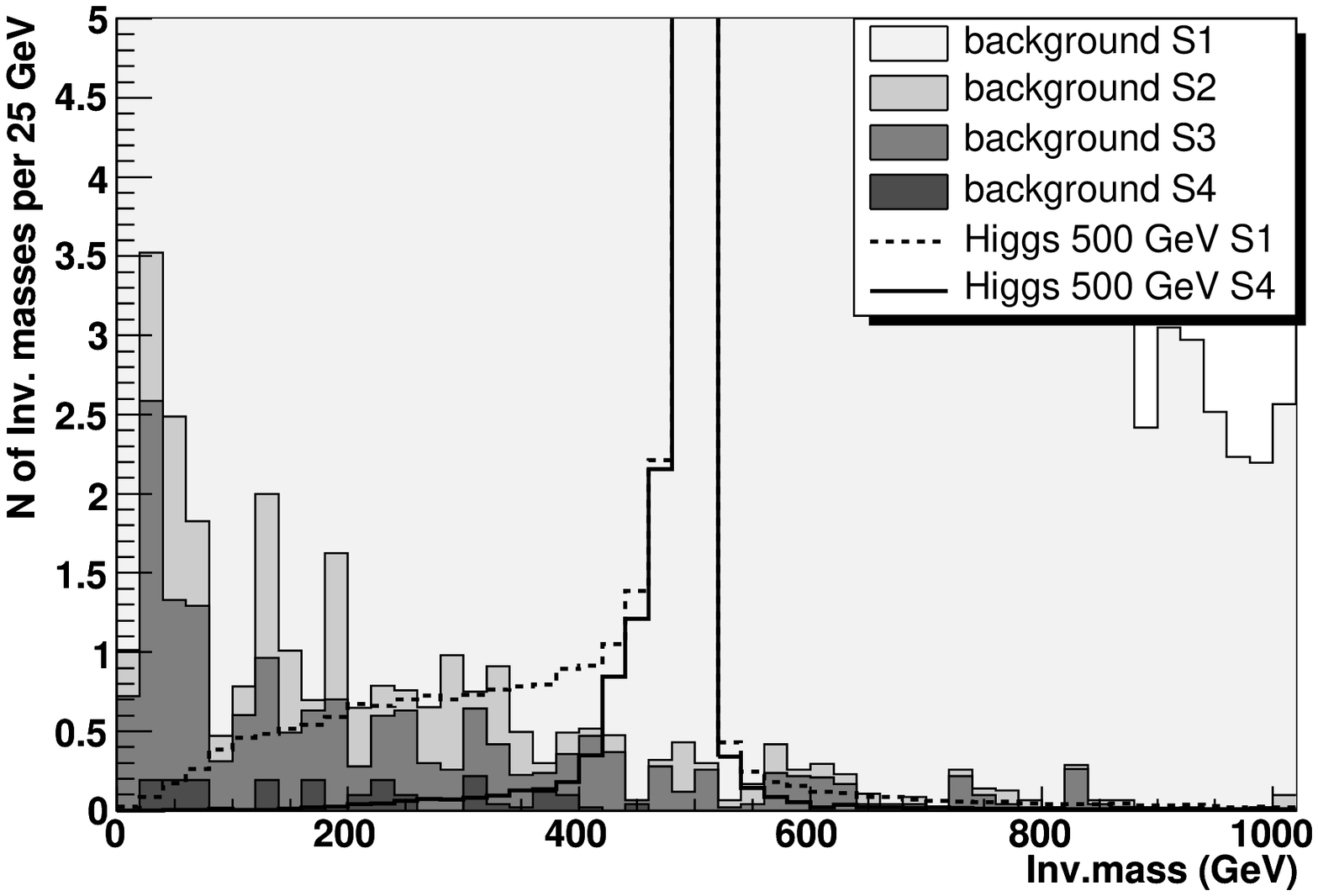}
\caption{Distribution of invariant masses after applying selection rules (S1-S4) for Higgs $M_{\Phi} = 500$ GeV and the Standard Model background (L=30 fb$^{-1}$). The histogram in the right panel is a zoom of the left histogram to illustrate the effects of the selection rules S2-S4.}
\label{f3}
\end{center}
\end{figure}

An example of invariant mass distribution after applying selection rules are shown in 
Figure~\ref{f3} for $M_{\Phi} = 500$ GeV. A tabulated example is given for $M_{\Phi} = 200, 500, 800$ GeV in Table~\ref{t5} corresponding to the 
luminosity $L=30 \mathrm{fb^{-1}}$. The strength of S2 cut is clearly visible: almost no decrease in signal while the number of the background events descends close to final minimum value. A peculiar behavior of S4 --  reducing the background, while also increasing the signal in its peak -- is the effect of 
applying the $\tau \to \mu'$ correction method described above. 

\begin{table}[h!]
\begin{center}
\begin{tabular}{|l|l|l|l|l|l|}
\hline\hline
Process & \multicolumn{5}{c|}{N of invariant masses} \\\cline{2-6}
        & N of $\Phi$ & S1 & S2 & S3 & S4 \\
\hline
\multicolumn{6}{|c|}{\emph{Energy range 150...250} GeV} \\
\hline
$M_\Phi$=200 GeV & 4670 & 1534 & 1488 & 1465 & 1539 \\
$t \bar t \to 4\ell$ & - & 1222 (168) & 172 (8.5) & 134 (6.9) & 17.6 (3.7) \\
$t \bar t Z$ & - & 21.3 (4.0) & 15.5 (1.0) & 6.3 (1.2) & 2.2 (1.1) \\
$ZZ$ & - & 95.0 (12.0) & 22.5 (0.7) & 9.8 (0.5) & 1.7 (0.2) \\
\hline
\multicolumn{6}{|c|}{\emph{Energy range 375...625} GeV} \\
\hline
$M_\Phi$=500 GeV & 119.2 & 48.4 & 47.5 & 46.8 & 49.5 \\
$t \bar t\to 4\ell$ & - & 178 (28) & 2.1 (0.9) & 1.65 (0.87) & 0.10 (0.35) \\
$t \bar t Z$ & - & 6.6 (1.7) & 2.3 (1.0) & 1.0 (1.0) & 0.00 (0.1) \\
$ZZ$ & - & 9.4 (2.9) & 1.4 (0.2) & 0.68 (0.19) & 0.08 (0.09) \\
\hline
\multicolumn{6}{|c|}{\emph{Energy range 600...1000} GeV} \\
\hline
$M_\Phi$=800 GeV & 11.67 & 5.05 & 5.00 & 4.92 & 5.21 \\
$t \bar t\to 4\ell$ & - & 77 (12) & 0.00 (0.22) & 0.00 (0.22) & 0.00 (0.07) \\
$t \bar t Z$ & - & 2.6 (1.2) & 0.39 (0.4) & 0.39 (0.4) & 0.00 (0.1) \\
$ZZ$ & - & 2.5 (0.8) & 0.34 (0.16) & 0.17 (0.09) & 0.00 (0.02) \\
\hline\hline
\end{tabular}
\end{center}
\caption{Effectiveness of the selection rules for the background and signal. All event 
numbers in the table are normalized for L=30 fb$^{-1}$. 
The numbers in brackets mark errors at 95\% confidence level for Poisson statistics. The signal increases after S4 due to the reconstructed $\tau \to \mu'$ decays.}
\label{t5}
\end{table}

\section{Including measurement errors to Monte Carlo}

In this Section we make an attempt to estimate  simplified detector effects at the
level of Monte Carlo analyses. In order to do that 
we have added overall detection efficiencies for the Monte Carlo generated $\mu$-s and $\tau$-jets -- 0.98 and 0.6 respectively. 
Additionally, we applied Gaussian distortion functions to Monte Carlo produced data 
for $\mu$-s, $\tau$-jets and $E_T^{miss}$ which  were used to alter randomly 
those quantities in the analysis. 
Although the precision of $\mu$ detection is sensitive to $p_T$ of $\mu$ and $|\eta|$ we use the mean values for a rough estimation. We make the following assumptions based on \cite{PDTRv2:2006a1,CMS:1997ki,PDTRv1:2006a3}. 

The direction of muon ($\tau$-jet) is altered with the Gaussian distribution: mean $\mu= 0.0005$ (0.031) and variance $\sigma^2 = 0.003$ (0.017). The transverse momentum is altered according to the $p_{T,\mathrm{rec}}/p_{T,\mathrm{Monte Carlo}}$ Gaussian distribution: mean $\mu= 1.$ (0.897) and variance $\sigma^2 = 0.03$ (0.089). Both components of missing transverse energy are altered independently according to the Gaussian distribution (mean $\mu= 0$ GeV and variance $\sigma^2 = 25$ GeV) by adding the piece to its Monte Carlo value.

The result of such a distortion is a decrease in both signal and background approximately by factor two (Table~\ref{t6}). As the background and the signal decrease proportionally, the luminosity needed for discovery roughly doubles. Remarkably the optimized S2 cut value does not change significantly. The proportion of the reconstruction channels in the total analysis has changed remarkably as shown in Table~\ref{t7}. The reason is clearly the small detection efficiency of $\tau$-jets. The $1\mu3j$ channel comprises only $1\% $ of the total signal if the detector effects are considered, while in the pure Monte Carlo analysis it forms $9\%$. The additional possible detector effects make the $\tau \to \mu$ correction even more relevant.

\begin{table}
\begin{center}
\begin{tabular}{|l|l|l|l|l|l|l|l|l|l|}
\hline\hline
Mass of $\Phi$ (GeV) & 200 & 300 & 400 & 500 & 600 & 700 & 800 & 900 & 1000 \\
\hline
Optimal $\sum p_T$ for S2 (GeV)      & 300 & 400 & 600 & 700 & 860 & 860 & 860 & 860 & 860 \\

Det Eff $L_{min}$ (fb$^{-1}$)  & 0.526 & 1.20 & 3.0 & 6.6 & 15 & 30 & 60 & 111 & 200 \\
Det Eff $L_{max}$ (fb$^{-1}$)  &0.546 & 2.19 & 6.5 & 16.6 & 39 & 86 & 190 & 420 & 900 \\
\hline\hline
\end{tabular}
\end{center}
\caption{Optimal $\sum p_T$ cuts and minimal discovery luminosities for different Higgs masses when the estimates of  detector measurement errors are taken into account. 
Two boundaries for the minimal luminosity are given: the lower ($L_{min}$) corresponds to the generated background and the higher ($L_{max}$) corresponds to 95$\%$ upper limit of the background error.}
\label{t6}
\end{table}

\begin{table}[h!]
\begin{center}
\begin{tabular}{|l|l|l|l|l|l|}
\hline\hline
 & \multicolumn{5}{c|}{Percentage of the channel after reconstruction} \\
\cline{2-6}
  Decay channel         & $4\mu$ & $3\mu 1j$ & $2\mu 2j$ & $1\mu3j$ & $\tau \to \mu$ correction \\
\hline
Monte Carlo & 21 & 28 & 26 & 9 & 16 \\
MC+ efficiencies   &  38 & 25 & 12 & 1 & 24 \\
\hline\hline
\end{tabular}
\end{center}
\caption{The importance of reconstruction channels at Monte Carlo level 
and considering detector efficiency effects.}
\label{t7}
\end{table}

\section{Conclusions and outlook}

We have studied possible direct test of little Higgs scenarios  which light  particle spectrum includes a triplet scalar multiplet  at LHC experiments. We have investigated the Drell-Yan
pair production of the doubly charged Higgs boson and its subsequent leptonic decays. 
In addition to solving the little hierarchy problem, this scenario can also explain 
the origin of non-zero neutrino  masses and mixing via the triplet Higgs neutrino 
mass mechanism. Simple connection between the observed neutrino mixing and triplet Yukawa 
couplings allows us to predict the leptonic branching ratios of the triplet. Thus the
experimental signatures of the model  do not depend on the size of the triplet Yukawa couplings
allowing direct tests of this scenario at LHC.

In our analyses we have considered  four $\mu$ and/or $\tau$ final states including up to 3 tau leptons.
 We propose four selection rules  to achieve the optimized  signal and background ratio. 
 As the $\Phi^{++}$ decays are lepton number violating, we have shown that the background can be 
 practically eliminated. In such an unusual  situation 
 we have used the LLR statistical method to demand $5\sigma$ discovery potential 
 to be bigger than 95\% ($1-\text{CL}_{s+b} > 0.95$) as the discovery criterion. The results of optimized cut values are presented in Table~\ref{t4}. Considering the pure Monte Carlo study, 
 $\Phi^{++}$ up to the mass 300 GeV can be discovered in the first year of 
 LHC ($L=1$ fb$^{-1}$) and $\Phi^{++}$ up to the mass 800 GeV can be discovered for the integrated luminosity $L=30$ fb$^{-1}$. 
Including particle reconstruction efficiencies as well as Gaussian distortion functions
for the particle momentas and missing energy which mimic detector inefficiencies at
Monte Carlo level, our results show that $\Phi^{++}$ can be discovered up to the 
mass 250 GeV in the first year of LHC and 700 GeV mass is reachable for the 
integrated luminosity $L=30$ fb$^{-1}$.

For further studies of this scenario at LHC 
progress can be made both physics-wise as well as technically.
Full simulations of the detector effects are needed which also include 
the electron, muon and tau  final states. 
For better determination of statistical errors coming from the 
background studies bigger 
SM background datasets must be produced. This requires huge computing resources.
If these goals can be achieved, the proposed phenomenology opens a new window 
to study the neutrino properties at colliders. In addition to the considerations in this paper,
one can determine at LHC experiments 
the hierarchy (normal or inverse) of light neutrino mass spectrum, 
and to estimate the two Majorana phases which are not measurable in neutrino oscillation
experiments \cite{meie}.

\vskip 0.5in
\vbox{
\noindent{ {\bf Acknowledgments} } \\
\noindent  
We thank the Estonian Science Fondation for the grant no. 6140. 
The work described in this paper was also supported by the European Union
through the FP6-2004-Infrastructures-6- contract No 026715 project
"BalticGrid".
}

\end{document}